\documentclass[sigconf, nonacm]{acmart}

\usepackage{pvldb}

\renewcommand\vldbdoi{XX.XX/XXX.XX}
\renewcommand\vldbpages{XXX-XXX}
\renewcommand\vldbavailabilityurl{https://gitlab.com/xdata/lasso} 

\usepackage{xcolor}
\usepackage{hyperref}
\hypersetup{
colorlinks=true, urlcolor=blue
}

\usepackage{booktabs}
\usepackage{longtable}
\usepackage{tabularx}
\usepackage{listings}
\lstdefinestyle{codestyle}{
  basicstyle=\ttfamily\scriptsize,
  breaklines=true,
  columns=fullflexible,
  frame=single,
  framerule=0.5pt,
  rulecolor=\color{black!20},
  backgroundcolor=\color{black!2},
  xleftmargin=4pt,
  xrightmargin=4pt,
  aboveskip=4pt,
  belowskip=4pt,
  captionpos=t
}
\usepackage{float}
\usepackage{algorithm}
\usepackage{algpseudocode}
\usepackage{tikz}
\usetikzlibrary{positioning}
\usepackage{pgfplots}

\pgfplotsset{compat=1.17}
\definecolor{colorEX}{HTML}{7F7F7F}
\definecolor{colorSpotIT}{HTML}{3CB44B}
\definecolor{colorSpotITVal}{HTML}{2E8B57}
\definecolor{colorPolygon}{HTML}{911EB4}
\definecolor{colorPolygonVal}{HTML}{BA55D3}
\definecolor{colorParSEval}{HTML}{FF8C00}
\definecolor{colorLasso}{HTML}{E6194B}
\definecolor{colorLassoCC}{HTML}{4363D8}
\definecolor{colorXDataAC}{HTML}{F58231}

\newcommand{\eat}[1]{}

\begin{document}

\title{LASSO: An LLM-Agent Synthesized System for SQL Validation}
\author{Shreyas Jadhav}
\affiliation{%
  \institution{IIT Bombay}
  \country{India}
}
\email{shreyasj@cse.iitb.ac.in}

\author{S. Sudarshan}
\affiliation{%
  \institution{IIT Bombay}
  \country{IIT Bombay, India}
}
\email{sudarsha@cse.iitb.ac.in}

\begin{abstract}
The rise of Large Language Models (LLMs) has significantly advanced text-to-SQL capabilities, but generated queries may have errors. 
Test datasets are important for catching errors, but having only one or a few test datasets leaves many bugs undetected.
Effective testing requires multiple test datasets that can expose subtle semantic bugs and edge cases.    
Recent work on test data generation and equivalence checking has focused on using constraint solvers.
However, constraint-based approaches are slow and not scalable.


In this paper, we introduce LASSO, a framework inspired by XData, but designed to minimize the use of constraint solvers, instead prioritizing use of data generation strategies based on fast heuristics.
LASSO deploys a number of ``attack strategies'', each targeting a specific class of semantic errors.

LASSO code was built using LLM-based agentic coding.  
We used carefully designed workflows and rules to support effective code generation and to avoid common code generation issues.
A large number of internal test cases are included in LASSO to catch errors in generated code, and more are added when we extend LASSO, to avoid subsequent regressions.
Our empirical evaluation across multiple benchmarks 
demonstrates that LASSO outperforms the state of the art, including XData, 
SpotIT, ParSEval, and Polygon.
\end{abstract}

\maketitle

\vldbtopmatter

\section{Introduction}

While Large Language Models (LLMs) can generate complex SQL queries from natural language, verifying the correctness of SQL queries remains challenging. Evaluating a query on a single or a few datasets (database instances) is often insufficient, as an incorrect query can coincidentally produce the expected output on those datasets.  Validation of correctness requires multiple test datasets  designed to expose subtle semantic bugs and corner cases.

Early data generators populated tables using only schemas and statistical distributions, in a query-independent fashion~\cite{gray1994quickly, bruno2005flexible, chays2004agenda}.  However, these
approaches, as well as query-independent fuzzing-based approaches, do not provide any guarantees related to
catching errors, or even generating non-empty results, on specific queries.
Reverse Query Processing (RQP)~\cite{binnig2007reverse}
and QAGen~\cite{binnig2007qagen} 
provided a symbolic approach to generate datasets on which query results satisfied desired properties.
Qex~\cite{veanes2010qex} translated SQL predicates into constraints, which are processed by an SMT constraint solver (Z3) to generate a dataset on which the query has a non-empty result.  None of these approaches addressed catching of errors.  

The  XData framework~\cite{gupta2010xdata, xdata2011, chandra2015data, Somwase2025TestDG} targeted dataset generation to catch errors in queries, and was shown to work well on moderately complex queries.
The current XData system generates multiple datasets to catch different types of errors, by generating appropriate constraint sets for each dataset, and solving each constraint set using an SMT solver.
Recent work on query equivalence refutation such as
\cite{verieql,Tremante2026SpotItVT,10.1145/3729303,Chen2025ParSEvalPT} is also based on constraint generation.
Fuzzing-based approaches such as \cite{zhong-etal-2020-semantic} are inefficient and less effective in catching errors with complex queries.   



We observed two main limitations of the existing work on SQL query validation: (1) scalability of time taken with increasing query complexity as well as variety of query features, and (2) efficacy at catching errors.
Approaches based on modeling queries as constraints which are then solved using SMT solvers are not very scalable to large query sizes.  Further each data type has to be separately handled.   Efficacy at catching errors is also limited by human resource availability to model complex query constructs and data types, and to code different approaches for data generation to catch errors.

While LLMs can be used to directly generate test cases, e.g.,~\cite{zhao2025llmsqlsolver}, direct LLM dataset synthesis incurs significant per-query latency and API token costs, as shown in our performance evaluation.

In this paper, we explore the idea of using an LLM to create a system for generating test data for SQL validation that can handle a wide variety of SQL query features, as well as a  wide variety of errors.  
Our goals and contributions include the following:
\begin{itemize}
\item Our first goal was to see if using agentic software development can generate a system that can perform better than existing systems, proving its effectiveness on a wide variety of benchmark schemas and queries. 

In Section~\ref{sec:lasso_overview} we describe how we built LASSO, an evolutionary framework heavily inspired by XData, designed to effectively leverage LLM-based agentic coding given an architectural setup, specific coding rules, and workflow instructions. 

LASSO comprises a collection of small imperative functions, called ``Attack Strategies'', each targeting a specific class of semantic mutations by detecting structural patterns in a given query and synthesizing datasets designed to kill the corresponding mutants.  

The attack strategy approach of LASSO makes it easy to handle new mutation types as they are discovered, by using an LLM agent to add new strategies or modify an existing strategy.

\item Our second goal was to explore efficient execution, by minimizing use of expensive SMT solver calls.  There is a tradeoff between the generality of relatively slow SMT solvers, and the efficiency of case-specific data generation approaches that are an alternative. 
Approaches that do not use SMT solvers, like Agenda~\cite{chays2004agenda} and the first version of XData \cite{gupta2010xdata} implemented several case-specific techniques such as adding/deleting tuples to ensure non-matching tuples for joins.  
However, they were later abandoned in favor of constraint-based approaches, which promise greater generality.

We wanted to see if despite avoiding SMT solver usage to a large extent, we can still create effective test cases by agentically creating a large number of case-specific approaches.   

To create a large pool of case-specific approaches, we relied on creating a large initial query test suite (details are provided in Section~\ref{sec:test_infra}), and iteratively improving our code by prompting the LLM to add code to handle queries from a larger benchmark that were not initially handled.   

There were two sub challenges in this task. 
\begin{itemize}
\item 
The first, to avoid regression as the code was iteratively improved, was handled by accepting changes only if they did not break existing successfully handled cases.  
\item The second challenge was to ensure that the agent did not hard-code solutions that would not generalize beyond the specific query.  We used a variety of instructions in the prompts/workflows to tell the agent to avoid such solutions.  We also performed code walkthroughs to understand what was  being implemented.
We also used LLMs to generate documentation from code, which we could check manually. 
We rejected changes where we detected hard-coding (although we may have missed some cases). 
\end{itemize}

\item 
In Section~\ref{sec:observations} we  report on observations from a review of the generated code.
Section~\ref{sec:base_data_generation} reports our observations on how base data (to generate non-empty results for a given query) is generated.
Section~\ref{sec:attack_strategies} describes observations on attack strategy generation, after summarizing the initial attack strategies we provided to the agent.

\item Our third goal was to create a comprehensive evaluation benchmark combining queries from existing benchmarks. We observed some limitations of existing approaches, including problems introduced by non-deterministic queries.  

We describe our evaluation setup in Section~\ref{sec:exec_test_setup}, including a normalization scheme that rewrites existing  benchmark queries, by modifying non-deterministic constructs to make them deterministic and rewriting queries to make them SQLite compatible (Section~\ref{sec:normalization}), and an execution result comparison setup (Section~\ref{sec:result_comp}).

\item Our fourth goal was to carry out a detailed performance evaluation of our agentically developed system LASSO. 

Section~\ref{sec:performance} describes our performance evaluation.
We created a unified benchmark, as described in Section~\ref{sec:unified_benchmark}, which has $36,042$ sanitized query pairs collected from eight different benchmark suites (BIRD~\cite{bird}, Calcite~\cite{10.1145/3729303}, Cubes \cite{10.1145/3729303}, LeetCode~\cite{10.1145/3729303, Tremante2026SpotItVT}, Literature~\cite{10.1145/3729303}, XData\-BM~\cite{xdata2011, Somwase2025TestDG}, TPC-H~\cite{Somwase2025TestDG}, and LargeJoin~\cite{Somwase2025TestDG}).

Our performance results (Section~\ref{sec:eval_results}) 
show that LASSO not only handles significantly more cases, showing non-equivalence, than other approaches, but it is also significantly faster than state-of-the-art tools, including SpotIT, ParSEval, and Polygon, while improving significantly on runtimes as compared to the alternatives.
 
This validates our original hypothesis that it is possible to build an effective and efficient 
test data generation system for SQL query validation, agentically.  

\item Our final goal was to document the development process to understand what approaches worked, which may help in other similar agentic development processes.  Our description of how LASSO was built (Section~\ref{sec:lasso_overview})
and our observations on generated code (Section~\ref{sec:observations}) provide some insight into this process.  We believe that ensuring we had a large number of initial test cases, coupled with appropriate prompts and workflows, and iterative improvement on further test cases all contributed to the success of our agentic code generation process.  
\end{itemize}

Our repository, \url{https://gitlab.com/xdata/lasso}, contains source code, prompts/workflow instructions, test cases and execution testing modules, scripts to run the experiments, and results.

 

\section{Background and Problem Statement}
\label{sec:background}
\label{sec:xdata}

Validating the semantic correctness of SQL queries is a challenge when writing complex SQL queries, for grading of student queries, and for evaluation of modern AI-driven text-to-SQL systems. Validating query correctness requires executing and evaluating the query on datasets that can distinguish subtle semantic differences between potential query variants.  A single static database is often insufficient for this purpose: an incorrect query may return the expected output on the given test database instance, although it may return different results on other database instances.
Consequently, validation of SQL queries requires generating multiple test datasets targeted to uncover potential semantic flaws (see, e.g., \cite{xdata2011}).

In practice, test data generation for SQL queries is employed in two key scenarios:
\begin{enumerate}
    \item \textbf{Standalone Query Testing:} A user (manually or using an LLM) writes a query $Q$ for a given task that is expressed in natural language, without a reference query. Multiple datasets can be generated from $Q$, to expose corner cases, such as boundary conditions, or null value handling, etc.  Checking the query result against the expected result on each dataset helps test the correctness of the query. 
    \item \textbf{Equivalence Refutation:} Scenarios where a query pair is available, such as in text-to-SQL evaluation, a candidate predicted query $Q_{\text{pred}}$ is evaluated against a reference query $Q_{\text{gold}}$. The goal is to synthesize a counterexample dataset such that the execution results of the queries differ, proving non-equivalence.  Recent work on query equivalence checking/refutation includes
    SpotIT~\cite{Tremante2026SpotItVT}, Polygon~\cite{10.1145/3729303}, and ParSEval~\cite{Chen2025ParSEvalPT}.
    These were shown to perform better than earlier generation systems such as Cosette \cite{cosette}, SPEC \cite{spes}, QED \cite{QED}, and VeriEQL \cite{verieql} in terms of query features handled.
\end{enumerate}
Our focus is on stand-alone query testing, although the test datasets generated by our approach can also be used for equivalence refutation.

\subsection{Problem Formulation and Mutation Testing}

In this section we give a formal problem statement, based e.g. on \cite{xdata2011}.
Given a database schema $\mathcal{S}$ and a SQL query $Q$, the result of executing $Q$ on a database instance $D$ is denoted by $Q(D)$.

A \textbf{mutation} of a query $Q$ is a syntactically valid variant query $Q'$ representing a potential semantic error, for example, replacing a comparison operator ($A < 5 \to A \le 5$), altering a join type (e.g., changing an \texttt{INNER JOIN} to a \texttt{LEFT OUTER JOIN}), or not considering NULL values when they should have been considered.  We refer to the variant query $Q'$ as a \textbf{mutant}. While syntactically different, some mutations may be semantically equivalent to $Q$; non-equivalent mutations, however, represent genuine semantic bugs.

A test dataset $D$ is said to \textbf{kill a mutation} (or \textbf{kill a mutant} $Q'$) with respect to query $Q$ if executing $Q$ and $Q'$ on $D$ produces divergent results, i.e., $Q(D) \neq Q'(D)$.  
A test suite with multiple datasets is said to kill a mutant $Q'$ of query $Q$ if the 
results differ on at least one of the datasets.

Given a query $Q$ and schema $\mathcal{S}$, the goal of single-query test data generation is to construct a compact suite of test datasets $\mathcal{D} = \{D_1, D_2, \dots, D_m\}$ that (ideally) kills all non-equivalent mutants of $Q$. In the context of equivalence refutation for a query pair $(Q_{\text{gold}}, Q_{\text{pred}})$, the objective is to synthesize at least one counterexample dataset $D \in \mathcal{D}$ such that $Q_{\text{gold}}(D) \neq Q_{\text{pred}}(D)$, refuting their equivalence.

We note that SQL permits certain non-deterministic constructs that can result in different executions of  the same or equivalent queries generating different results.
Examples of such constructs include the use of the SQL LIMIT clause, and the use of time-dependent functions such as current date/time.   Such queries can be made deterministic by rewriting the query, for example by having an ORDER BY clause that includes all output attributes whereever LIMIT is used, and replacing calls to time-dependent and random-value generation functions by functions returning fixed values. 

We assume in the rest of our description that queries provide deterministic answers. 


\paragraph{Terminology.}
In different contexts, different terms are used to describe queries and their variants. In this paper:
\begin{itemize}
    \item We use \textbf{Reference Query} and \textbf{Gold Query} interchangeably to denote the query considered to be correct ($Q$ or $Q_{\text{gold}}$).
    \item We use \textbf{Candidate Query}, \textbf{Prediction Query}, and \textbf{Mutation Query} (or \textbf{Mutant}) interchangeably to denote the variant query ($Q'$ or $Q_{\text{pred}}$) whose semantic equivalence to the reference query is being evaluated. In practice, such queries may be generated by text-to-SQL models, written manually (e.g., by students), or produced by other tools such as query optimizers or mutation generators.
\end{itemize}

\subsection{The XData Approach}

Since LASSO is based on the single-query test data generation methodology of XData~\cite{gupta2010xdata, xdata2011, chandra2015data, Somwase2025TestDG}, we briefly outline the XData methodology below. Rather than requiring a query pair, XData utilizes a single query $Q$ and its schema $\mathcal{S}$ to generate a suite of datasets $\mathcal{D}$.

The methodology for single-query generation in XData follows a two-stage process:
\begin{enumerate}
    \item \textbf{Baseline Non-Empty Dataset:} The system first generates a baseline dataset $D_{\text{base}}$ designed to produce a non-empty result for the reference query ($Q(D_{\text{base}}) \neq \emptyset$). This baseline dataset immediately kills all mutations that evaluate to an empty result on $D_{\text{base}}$.
    \item \textbf{Targeted Mutation-Killing Datasets:} The system subsequently derives additional datasets specifically crafted to kill mutations across a targeted mutation space (such as comparison boundaries, join types, NULL handling, aggregates, and subqueries).
\end{enumerate}

This methodology has evolved across two main paradigms:
\begin{itemize}
    \item \textbf{Imperative and Algorithmic Generation:} The initial version of XData~\cite{xdata2011} achieved this imperatively and algorithmically. It operated by first synthesizing a base dataset that satisfied the original query. It then systematically applied imperative modifications to this baseline, such as selectively omitting certain tuples or altering attribute values to kill specific mutants, while preserving database integrity constraints.

    \item \textbf{SMT-Based Constraint Solving:} Subsequent versions of XData~\cite{xdata2011,chandra2015data,Somwase2025TestDG} accomplish this mutation-killing process by generating and solving constraints using an SMT solver (such as Z3). This approach models tables as arrays of tuples with constraint variables, generates constraints between the variables based on integrity and query constraints, and utilizes an SMT solver (such as Z3) to generate datasets satisfying the constraints. 
    
    One dataset is created using constraints designed to generate a non-empty query result. To kill a specific mutation, the system modifies the constraints at the targeted relational algebra node,  by ensuring that the unmutated node yields a result that diverges from the mutated node.

\end{itemize}

While this SMT-driven approach provides formal mathematical guarantees, it can become computationally intensive as query complexity and the required number of tuples increase. Because relations and intermediate results are modeled as bounded arrays of symbolic tuples, increasing the maximum number of tuples introduces additional constraint variables and constraint predicates for each tuple. Consequently, synthesizing datasets for complex queries or scenarios requiring larger tuple bounds places a significant burden on the solver, often leading to performance overhead and timeouts. 

\section{The LASSO Framework}
\label{sec:lasso_overview}

In this paper we present the LASSO framework, which is an extension of the XData approach.
Like XData, LASSO  retains a similar high-level methodology: given a query, generate a base data designed to provide non-empty results, and then generate multiple targeted datasets to kill mutations.
Each such dataset is created using one approach from a collection of approaches that we call \textit{Attack Strategies}. Revisiting the initial XData idea from~\cite{xdata2011}, LASSO generates datasets algorithmically
for the most part, minimizing reliance on SMT solvers. 

LASSO was built using agentic coding, and designed to support a wide variety of SQL queries. Human input to the agents included the codebase architecture, pre-defined workflows and instructions, and a testing setup.

We used Google Antigravity for code development, with Gemini (versions 3.1 to 3.8) as the primary LLM, but also used Claude Opus 4.6 for some tasks.  We used a Gemini AI Pro subscription over a 5 month period.
The overall code development was done by just one person, and approx.\ 10k lines of code were generated.

LASSO uses a design approach of having separate attack strategies of each type of semantic bug, which are stored in separate files named based on the error being targeted.  When an agent needs to modify code to handle a semantic error that is not currently caught, this approach makes it easy for the agent to navigate to the exact attack strategy files, and load only the required relatively small attack strategy code into context and work in isolation without affecting other attack strategies. This also makes it easy to support new type of bugs by creating corresponding attack strategy with  applicability rules for the strategy, instead of having to work on a central large strongly-connected code base.  A central large algorithm also means that an LLM agent has to load longer input context, which can result in performance degradation \cite{liu-etal-2024-lost}. 


\subsection{Setup for System Generation}

The basic setup for generating the system consisted of two parts: an initial architecture, and a testing setup.  We summarize the prompts used to create these two parts in this section.   
To generate meaningful data, we used the Faker library \url{https://github.com/joke2k/faker}.


We provided an initial scaffolding prompt explaining the core idea of the project. 
The prompt asked the coding agent to implement a basic architecture, by  providing it a prompt for each component of the architecture.  The prompts are summarized below.

\textbf{Parsing Modules Prompt:}  ``The DDL should be parsed using SQLGlot and converted into a Schema representation containing tables, columns, data types, primary keys, UNIQUE constraints, NULLability, foreign keys, and CHECK constraints. 
    \begin{itemize}
        \item While parsing columns, include metadata such as corresponding Faker domain, determined using heuristics such pattern matching on column names. (e.g., first\_name to person-name generators, address to location/address generators, company to organization names). This metadata should be later used to come up with reasonably realistic looking domain-specific values using the Faker library.''
    \end{itemize}

\textbf{Base Dataset Generator Prompt:} ``The base dataset generator module should be responsible to construct a base dataset that would produce non-empty results for the given query and schema.''

\textbf{Dataset Post-processing Prompt:} ``After dataset generation, this post-processing stage should ensure consistency. This includes repairing potential foreign-key violations, etc. to avoid issues with loading the dataset and remain valid under the declared schema.''
        
\textbf{Attack Strategies Prompt:} ``Each attack strategy should focus on a specific class of semantic mutations (for example, inverted comparisons, omitted \texttt{DISTINCT}, or improper join types). Each attack strategy should be placed separately as a module in a separate file in the \texttt{/attacks} directory. All strategies should implement a common \texttt{AttackStrategy} interface with two core operations:
        \begin{itemize}
            \item \textbf{Is Applicable:} Analyzes the parsed query AST to detect the presence of query features or constructs where the specific semantic error (mutation) targeted by the strategy can potentially occur.
            \item \textbf{Generator:} Constructs one or more \texttt{Dataset} objects specifically designed to kill the corresponding mutant queries. Each generated dataset should contain the synthesized data, the strategy name, tags, and a description of the targeted mutation and dataset design.''
        \end{itemize}
We describe later the approach for creating datasets for each attack strategy that was used by  the agent.

\textbf{SQL Serialization Prompt:} ``The internal relational representation will be converted into executable SQL (INSERT INTO...) statements. Tables should be serialized in dependency order so that referenced tables are inserted before dependent ones. Values should be formatted appropriately for SQL types, including NULL handling, string escaping, etc.''

\textbf{Entry Point Prompt:} ``The main entry point should receive the DDL and SQL query, constructs schema and query representations. It should then execute the applicable attack strategies. Each strategy should produce one or more datasets, which are post-processed, serialized, and finally de-duplicated.''

\subsection{\textbf{Testing Infrastructure}} ~
\label{sec:test_infra}

The success of agentic coding depends on having a strong test suite.  We therefore created a large test suite as described below, and also created test utilities to verify that the generated datasets were catching errors.   

When we subsequently found test cases which were not caught by our code, we added these to the test suite and prompted the agent to create appropriate attack strategies for the new test cases.


\begin{itemize}
    \item \textbf{Test Cases:}\label{sec:initial_test_cases} To bootstrap development, we constructed initial test cases using the semantic error descriptions from \textit{NL2SQL-BUGs} by Liu et al.~\cite{Liu2025NL2SQLBUGsAB}. The paper provides a detailed taxonomy and descriptions of SQL semantic errors categorized into 9 main categories and 31 subcategories. We passed the content to an LLM, prompting it to generate a representative multiple test cases for each subcategory. Each test case is a JSON file comprising the following properties:
    \begin{itemize}
        \item A database schema DDL
        \item A reference (gold) query passed for data generation
        \item Expected Mutants to be Killed: An array of candidate queries representing semantic mutations (mutants). Each mutant query is associated with a description of the targeted mutation.
    \end{itemize}
    \item \textbf{Execution Tester:} A separate module that executes both the reference query and each candidate mutant on the generated datasets to detect whether a given dataset refutes the equivalence, along with feedback on the execution details. 
    \item \textbf{Regression Suite:} For each example that is successfully handled, a minimal test case is added to the regression suite, ensuring future modifications do not break previously supported functionality.
    \item \textbf{Module Testing:} Beyond execution testing, we checked that the core data generator module supports all test suite queries, and produces datasets yielding non-empty query results.
    Any failure results in a prompt to modify the code to properly handle the failed query.
    \item \textbf{Helper Scripts}: 
    \begin{itemize}
        \item \textbf{Run All:} A dedicated script that prints the only feedback that is required for an agent, after running all the test cases. Such as only the failing cases if any.
        \item \textbf{Improvements Finder:} A script that automatically gathers examples that current codebase does not handle. The examples are grouped into two categories: (1) \textbf{Other-Proved Non-Equivalent Queries:} Finds instances where other tools (such as \textit{ParSEval}, \textit{Polygon}, \textit{SpotIT}) successfully proved query non-equivalence but LASSO did not. Such confirmed non-equivalent queries are then used to improve the LASSO codebase.\\
(2)  \textbf{Unknown Queries:} Collects the remaining examples where none of the tools proved non-equivalence.  For such queries we ask the LLM to check for equivalence directly using its own reasoning, and  proceed to implement support in the codebase only if LLM determines non-equivalence.

    \end{itemize}

\end{itemize}

\subsection{Instructions and Workflows}

After the initial scaffolding prompt, we created multiple re-usable prompts and instructions to extend the framework. 

\subsubsection{\textbf{Rules}}~

Upon observation of coding agent's output and behavior over multiple iterations, we created a set of rules to guide the agent to avoid some common mistakes in future iterations and improve overall code generation quality through correct reasoning. We attached these rules to every prompt. 
Below, we explain the rules at a high level. (The full set of rules is available in our code repository.)

\textbf{Reasoning related rules} are designed to guide the LLM's reasoning to help in coming up with better quality solutions.  For example, explicitly asking the LLM to consider reasoning about SQL semantics, relational algebra, and to create generalized solutions.  The rules also instruct the LLM to first reason about exactly what type of data would be needed for a given case, what would be general steps to construct such datasets, reason whether the solution always work, or is any better choices available, and to check if the approach being considered would break existing features, etc.

\textbf{Implementation related rules} mention overall implementation style that the project should follow. Such as prioritizing programmatic, imperative algorithms over SMT solver formulations for dataset construction, while also outlining cases where it should consider using SMT solvers (e.g. complex expressions). 
We also include rules telling the agent to avoid common implementation mistakes, for example, rules to make sure each strategy focuses on a single category of semantic errors, data generation is deterministic and targeted, without relying on simple random data generation, avoiding hard-coding of fixed values or benchmark specific assumptions, avoiding cluttering with redundant logic, etc.

\textbf{Testing and debugging related rules} emphasize testing and debugging through the prescribed test runner, verifying that the generated dataset exposes the intended semantic difference, rather than an unrelated discrepancy, steps to inspect the results, avoiding getting stuck running long scripts, etc.

\subsubsection{\textbf{Workflow}}~

For any failing benchmark instance or test cases, we provided the agent with a set of steps to follow to address the failure, which helped in avoiding token wastage.
Below we highlight a few important workflow instructions.  (Detailed workflow instructions are available in our code repository.)
\begin{enumerate}
    \item For a given example query pair and a schema, first reason whether the queries are semantically non-equivalent. If the queries are semantically equivalent, notify the user and terminate immediately. 
    \item If the queries are semantically non-equivalent, modify only the src/ codebase to generate required test data for proving non-equivalence (NEQ) of query pairs in the example. If changes to testing scripts are required, first request a review from user.
    \item Before coding, create a minimal reproducible test case in the \texttt{test\_cases/} directory.
    \item Before implementation, verify directly (not generating by code) that the dataset would actually proves the queries are not equivalent.
    \item Implement a required generalized solution in \texttt{src/} 
    \item Verify the functionality of \textbf{base data generator} that it produces non-empty results for the reference query and verify generating data using the mutant query produces non-empty results for the mutant.
    \item Verify the minimal test case passes after the implementation. and finally, verify the original provided example also passes.
    \item  Ensure your solution did not cause any regressions, if there are any regression, you will re-consider your solution and fix them.

\end{enumerate}

\subsubsection{\textbf{Routine Checks}}~

After each iteration, we manually skimmed through the generated diffs to inspect code quality and integrity.  We looked for shortcuts such as modifying test runners, or hardcoding, etc. instead of implementing genuine data generation logic, and when we found it, we reverted the code changes, and later added instructions in the prompt to avoid recurrences.

\subsubsection{\textbf{Refactoring}}~

To maintain code quality and prevent architectural degradation across iterations, we conducted separate refactoring sessions. We provided LLM agents with refactoring rules, mainly to follow standard software engineering principles (e.g., separation of concerns, eliminating duplication, maintaining simplicity, ensuring observability, and preserving existing behavior).

In practice, coding agents occasionally hallucinated code or attempted shortcuts that violated core guidelines, such as generating random data, hardcoding values, or modifying testing scripts. As such issues were noticed, we updated the rules and workflows accordingly, to avoid those mistakes in future iterations. Combined with manual verification (especially for testing related code) and dedicated LLM-based routine code-reviews and refactoring sessions, the overall approach became effective for quickly supporting a large number of SQL features while maintaining system reliability.

\subsection{LLM-Human Interaction}
\label{sec:workflow_case_studies}

When implementing any code, the agent was asked to first suggest an implementation plan.  The human then had a dialog with the agent to discuss the plan in more detail, and to suggest changes in the plan.
Finally, the agent was asked to implement the code according to the mutually agreed plan.
If any issues were found later, the agent was asked to check what could be the root cause.  After suggestions on how the plan could be improved based on the causes found, the agent regenerated the relevant part of the code as per the refined plan.
Some examples of interaction can be found in our repository.

\section{Observations on Generated Code}
\label{sec:observations}

The agent came up with high level strategies as well as detailed implementation approaches on its own.  
We reviewed the core modules and attack strategies, to gain insights and understand how the system actually works. 
Below, we provide high-level explanations of the working of base data generation (Section~\ref{sec:base_data_generation}), and some of the attack strategies (Section~\ref{sec:attack_strategies}) from the generated code.\footnote{Detailed explanations generated by LLM-assistance (manually verified) are provided in our code repository.}
We also discuss below what could have been the literature/code source that inspired the agent to come up with the chosen approach.

\subsection{Base Data Generation}
\label{sec:base_data_generation}

The base dataset generation pipeline constructs datasets on which the given query produces a non-empty result. Below is an overview of the core sequential steps executed by the pipeline (details can be found in our code repository):

 \textbf{1. Equivalence Class Construction.}
    The code groups columns that must share identical values for the query to produce a non-empty result (e.g., join keys, foreign key pairs, aliased references to the same table column)  into equivalence classes using a Union-Find structure. 
    
    \emph{Relation to Existing Techniques:} Partitioning relational attributes into equivalence classes to capture join dependencies originates in query optimization and size estimation~\cite{swami1994estimation}, and was subsequently adopted by constraint-based test data generators such as XData~\cite{xdata2011, chandra2015data} to consolidate predicates across equi-joins.

 \textbf{2. Constraint Collection \& Cardinality Analysis.}
    For each equivalence class, all column constraints from the query and the schema on those columns are gathered into a single \textit{ColumnSpec} record. This includes allowed and disallowed values, range bounds, uniqueness constraints, \texttt{CHECK} expressions, and flags indicating whether column is part of certain clauses such as GROUP BY or sub-query, etc. This step also computes $\mathcal{K}_{\min}$, the minimum number of rows that may be needed for queries to return non-empty results. 
    
    Depending on the query complexity, the Z3 solver is used to estimate cardinality and to find seed values. However, the constraint formulations also make simple assumptions and use heuristics, some of which are sketched later in this section.
    
    The approach of collecting constraints is implicit in earlier work on data generation, and may have been present in the corresponding code that was used to train the LLM.

\textbf{3. Column-Centric Value Resolution.}
    Once constraints are aggregated and the target row count $\mathcal{K}$ is determined, this stage synthesizes concrete values for columns of each equivalence class root. 
    We observed that the code used several heuristics, and used SMT solver only in a few cases.
    
    Examples of heuristic choices include choosing a starting value and incrementing it to generate values for integer valued primary key/unique valued columns, and having different starting point for different columns to detect wrong column selection errors.
    
    For cases with constraints, heuristics included 
    choosing a midpoint for bounded range constraints, directly choosing literals from query for satisfying equality conditions, and using small SMT formulations and calling z3 for resolution of arithmetic expressions across multiple columns.
    
    Heuristic choices related to strings included generating strings by replacing wildcards in regex patterns with fixed values, trimming strings to satisfy length constraints, and using fake values from Faker library (keeping column names in mind) for unconstrained strings.
    
    Aggregation related heuristics included assuming all tuples are in a single group to simplify data generation, and generating pairs of values equally spaced from a desired average value, to ensure the desired average is generated.

    When constraints are violated, the code iterates over checking and heuristic repairs until valid assignments are found.  The agent is asked to update the code  when we find the code is not able to handle a given query. 

    We note that while there would certainly be some cases where heuristics fail, the performance results show that, overall, LASSO significantly outperforms other approaches in catching errors as well as in time taken. 
    The heuristics are straightforward, but we could not find any specific source from which the LLM could have learnt the heuristics.
 
\textbf{4. Row Assembly \& Cross-Column CHECK Repair.}
    Individual column value lists produced above are assembled into complete row records one column at a time.
    If certain value combination violates the check condition (e.g., \texttt{end\_date} was generated earlier than \texttt{start\_date}), the repair mechanism fix such violations heuristically as well (e.g., advancing \texttt{end\_date} past \texttt{start\_date}). Any adjusted candidate value is verified through the constraint checker against the dependent column's own ColumnSpec.

\textbf{5. Augmentation.} This step allows the base dataset to already detect mutations in outer joins (\texttt{LEFT JOIN} mutated to \texttt{INNER JOIN}) or anti-joins (\texttt{NOT IN} subqueries).  The pipeline appends deliberately unmatched parent rows. 

This strategy is somewhat similar to the approach of \cite{gupta2010xdata}, which deletes tuples to get a similar effect.  The strategy was observed to have been created in response to a prompt asking for minimization of datasets where possible.

\textbf{6. Post-Processing \& Serialization.}
A final consistency pass prepares the datasets for execution: (1)~\textbf{Orphan Row Repair} recursively creates parent rows for any child rows referencing missing foreign keys; (this high level idea was provided by a human via prompts, however the agent handles the implementation details such as transitively fixing foreign-key dependencies). (2)~\textbf{Uniqueness Enforcement} deduplicates primary and unique keys; (3)~\textbf{Topological DAG Serialization} writes out SQL \texttt{INSERT} statements in foreign-key dependency order; and (4)~\textbf{Dataset Deduplication} removes duplicate datasets across attack strategies via canonical query hashing.

We note that the above step is applied also to datasets created by attack strategies.

\subsection{Targeted Attack Strategies}
\label{sec:attack_strategies}

Attack strategies either parameterize the base data generator pipeline (such as by injecting specific values or nulls for specified columns, or omitting specified tables from the generated dataset) or attack strategies may mutate the generated base dataset in certain ways to create a test dataset (e.g. as duplicating a row or modifying individual values).

\subsubsection{\textbf{Initial manually provided Strategies}}
\label{sec:scaffolding_strategies}

As part of the initial scaffolding prompt 
we provided the LLM agents with examples of basic attack strategies (only ideas, not actual implementation). These initial strategy ideas established the basic design pattern of inspecting the parsed SQL AST for applicable syntactic patterns and ideas to synthesize datasets for killing various mutations. For example, the \textbf{NullHandlingAttackStrategy}  identifies nullable columns referenced in \texttt{IS NULL} checks and injects \texttt{NULL} values specifically into those columns.
The \textbf{DistinctAttackStrategy} targets queries missing a \texttt{DISTINCT} modifier. It generates datasets with intentionally duplicated rows in the referenced tables.
The \textbf{ComparisonAttackStrategy} targets comparison operator mutations (\texttt{>}, \texttt{<}, \texttt{=}, \texttt{BETWEEN}). It extracts all literal boundaries from the AST predicates and generates exact boundary values alongside slight offsets (e.g., $V$, $V+1$, $V-1$) using value injections. 

These seed strategies served as examples demonstrating how an attack strategy remains lightweight and imperative, avoiding SMT formulations.

\subsubsection{\textbf{Observations on Agent-Generated Strategies}}
\label{sec:generated_strategies}

The remaining attack strategies in LASSO were generated by the LLM agents to support our curated collection of initial test cases, created as described in Section~\ref{sec:initial_test_cases}. During iterative framework development, we initially aimed to support all of the failing test cases, one at a time. In response to supporting each failing test case, and taking hints from descriptions, the agent updated or created new targeted attack strategies. Some strategies may also have been created or evolved to support specific examples directly from existing benchmarks such as BIRD, XDataBM, etc.
Below, we detail our observations and key insights for a representative subset of attack strategies generated by the agent:

\begin{itemize}
\item \textbf{Join Attack Strategy:}
\label{sec:strategy_join}
To kill join type mutants (such as mutating an \texttt{INNER JOIN} to a \texttt{LEFT OUTER JOIN} or vice versa) this strategy works as follows:
\begin{enumerate}
    \item  Extract all distinct table aliases $\{T_1, T_2, \dots, T_k\}$ referenced in the query ($k > 1$).
    \item  For each alias $T_i$, invoke the base generator, specifying $T_i$ is to be suppressed. The generator suppresses tuple generation for table alias $T_i$, producing a dataset where $T_i$ contains zero rows while all other tables $T_{j \neq i}$ remain populated with valid records.
\end{enumerate}
The above strategy is directly based on
the strategy in \cite{gupta2010xdata}.

For example, to distinguish a \texttt{LEFT JOIN} from an incorrect \texttt{INNER JOIN}, the strategy generates a dataset that populates the left table (\texttt{Dept}) while leaving the right table (\texttt{Emp}) completely empty:
\begin{center}
\scriptsize
\begin{tabular}{cc}
  \begin{tabular}[t]{lc}
    \toprule
    \multicolumn{2}{c}{\textbf{Dept}} \\
    \texttt{id} & \texttt{name} \\
    \midrule
    1 & \textquotesingle HR\textquotesingle \\
    \bottomrule
  \end{tabular}
  &
  \begin{tabular}[t]{lc}
    \toprule
    \multicolumn{2}{c}{\textbf{Emp}} \\
    \texttt{d\_id} & \texttt{title} \\
    \midrule
    \multicolumn{2}{c}{\textit{(0 rows / empty)}} \\
    \bottomrule
  \end{tabular}
\end{tabular}
\end{center}
\eat{
(Under \texttt{LEFT JOIN}, the unmatched \texttt{Dept} tuple is retained with a \texttt{NULL}-extended title, yielding \texttt{[('HR', NULL)]} ($1$ row). An \texttt{INNER JOIN} requires matches from both tables, returning $0$ rows.)
}

\item \textbf{Aggregate Null Attack Strategy:}
\label{sec:strategy_agg_null}
To kill mutants with incorrect \texttt{NULL} handling in aggregate functions (such as replacing \texttt{COUNT(*)} with \texttt{COUNT(col)} or unintended \texttt{NULL} propagation in \texttt{SUM}, \texttt{AVG}, \texttt{MIN}, \texttt{MAX}), this strategy injects \texttt{NULL} values into nullable aggregate arguments:
\begin{enumerate}
    \item It traverses aggregate function nodes in the query AST to find non-primary-key, nullable column references $(T, C)$ (or all nullable columns for \texttt{COUNT(*)}).
    
    \item For each target $(T, C)$, the strategy synthesizes three datasets:
    
\textbf{(a.) Only-Nulls Configuration:} Generates two separate partitions—one standard group with valid non-null values, and a second group where every row has column $C$ set to \texttt{NULL}:
        \begin{center}
        \scriptsize
        \begin{tabular}{lccc}
          \toprule
          \multicolumn{4}{c}{\textbf{Emp}} \\
          \texttt{id} & \texttt{dept} & \texttt{bonus} & \textit{Partition} \\
          \midrule
          1 & \textquotesingle Eng\textquotesingle & 500 & \textit{Group 1 (Valid)} \\
          2 & \textquotesingle HR\textquotesingle & \texttt{NULL} & \textit{Group 2 (Only-Null)} \\
          3 & \textquotesingle HR\textquotesingle & \texttt{NULL} & \textit{Group 2 (Only-Null)} \\
          \bottomrule
        \end{tabular}
        \end{center}
        (Differentiates \texttt{COUNT(bonus)} from \texttt{COUNT(*)}. 
        \eat{
        On \textquotesingle HR\textquotesingle, \texttt{COUNT(bonus)} ignores \texttt{NULL} tuples and returns $0$, whereas \texttt{COUNT(*)} returns $2$.)
        }
        
\textbf{b. Mixed-Nulls Configuration:} Creates a single group containing both valid and \texttt{NULL} values for column $C$. It forces \texttt{NULL} rows and non-null rows to share identical \texttt{GROUP BY} keys (using \texttt{value\_injections}).
\eat{
        \begin{center}
        \scriptsize
        \begin{tabular}{lccc}
          \toprule
          \multicolumn{4}{c}{\textbf{Emp}} \\
          \texttt{id} & \texttt{dept} & \texttt{bonus} & \textit{Partition} \\
          \midrule
          1 & \textquotesingle HR\textquotesingle & 300 & \textit{Group 1 (Mixed)} \\
          2 & \textquotesingle HR\textquotesingle & \texttt{NULL} & \textit{Group 1 (Mixed)} \\
          3 & \textquotesingle HR\textquotesingle & 700 & \textit{Group 1 (Mixed)} \\
          \bottomrule
        \end{tabular}
        \end{center}
        (Exposes improper \texttt{NULL} propagation in \texttt{SUM(bonus)} or \texttt{AVG(bonus)}. Standard SQL skips \texttt{NULL} rows (computing $\text{AVG} = 500$ across $2$ valid rows), killing mutants that divide by the total row count ($N=3$) or evaluate the sum to \texttt{NULL}.)
}
        
\textbf{c. Standalone-Null Dataset:} Emits an isolated dataset containing strictly rows where column $C$ is \texttt{NULL} without any non-null groups:
        \begin{center}
        \scriptsize
        \begin{tabular}{lcc}
          \toprule
          \multicolumn{3}{c}{\textbf{Emp}} \\
          \texttt{id} & \texttt{dept} & \texttt{bonus} \\
          \midrule
          1 & \textquotesingle HR\textquotesingle & \texttt{NULL} \\
          2 & \textquotesingle HR\textquotesingle & \texttt{NULL} \\
          \bottomrule
        \end{tabular}
        \end{center}
        (Tests scalar aggregations and subqueries (e.g., \texttt{SELECT AVG(bonus) FROM Emp}), verifying that the aggregate correctly evaluates to \texttt{NULL} rather than defaulting to $0$.)

\end{enumerate}

\item \textbf{Group-By Attack Strategy:}
\label{sec:strategy_group_by}
To kill mutants related to \texttt{GROUP BY} clauses, this strategy works in three different ways:
\begin{enumerate}
    \item \textit{Hierarchical Prefix Synthesis (for queries with \texttt{GROUP BY}):} When a query groups by multiple columns (e.g., \texttt{dept} and \texttt{role}), a mutant query might accidentally drop trailing grouping columns (e.g., grouping only by \texttt{dept}). To catch this, the strategy generates rows that share the same leading grouping keys (using value injections) but have distinct values for subsequent grouping keys:
    \begin{center}
    \scriptsize
    \begin{tabular}{lcc}
      \toprule
      \multicolumn{3}{c}{\textbf{Emp}} \\
      \texttt{id} & \texttt{dept} & \texttt{role} \\
      \midrule
      1 & \textquotesingle Eng\textquotesingle & \textquotesingle Dev\textquotesingle \\
      2 & \textquotesingle Eng\textquotesingle & \textquotesingle QA\textquotesingle \\
      \bottomrule
    \end{tabular}
    \end{center}
    (A query grouping by both \texttt{(dept, role)} partitions these rows into $2$ distinct groups ($2$ output rows). A mutant query grouping only by \texttt{dept} collapses them into $1$ single group ($1$ output row).)

    \item \textit{Empty Relation Synthesis:} Synthesizes an empty dataset containing zero tuples ($0$ rows) across all tables:
    \begin{center}
    \scriptsize
    \begin{tabular}{c}
      \toprule
      \textbf{Emp} \\
      \midrule
      \textit{(0 rows / empty)} \\
      \bottomrule
    \end{tabular}
    \end{center}
    (In SQL, a scalar aggregate without \texttt{GROUP BY} over an empty table returns exactly $1$ row containing default or \texttt{NULL} values (\texttt{(0)}). Conversely, a query with an explicit \texttt{GROUP BY} over an empty table produces $0$ groups ($0$ rows). This output asymmetry kills mutants with missing or spurious \texttt{GROUP BY} clauses.)

    \item \textit{Multi-Group Synthesis for Scalar Aggregations:} For global scalar aggregate queries, a mutant might incorrectly introduce an unnecessary \texttt{GROUP BY} clause. The strategy generates multiple rows with distinct attribute values:
    \begin{center}
    \scriptsize
    \begin{tabular}{lcc}
      \toprule
      \multicolumn{3}{c}{\textbf{Emp}} \\
      \texttt{id} & \texttt{dept} & \texttt{salary} \\
      \midrule
      1 & \textquotesingle Eng\textquotesingle & 1000 \\
      2 & \textquotesingle HR\textquotesingle & 800 \\
      \bottomrule
    \end{tabular}
    \end{center}
    (The correct scalar aggregate combines all rows into $1$ global summary row, whereas a mutant that adds \texttt{GROUP BY dept} splits the output into multiple separate group rows ($2$ rows).)
\end{enumerate}

\item \textbf{Null in Subquery Attack Strategy:}
\label{sec:strategy_null_subquery}
In SQL, \texttt{WHERE col NOT IN (subquery)} type of queries may introduce subtle Three-Valued Logic (3VL) bugs: if the subquery returns even a single \texttt{NULL} value, the entire \texttt{NOT IN} condition evaluates to \texttt{UNKNOWN} (treated as false in \texttt{WHERE}), causing the query to return an empty result. This strategy synthesizes a dataset where the subquery explicitly contains a \texttt{NULL} value in the join column:

\begin{enumerate}
    \item  Traverses the query AST for \texttt{NOT IN} subqueries, identifies the projected subquery column, and verifies via the schema that the column is nullable.
    \item  Invokes the base generator to synthesize valid outer rows satisfying outer predicates and joins.
    \item Clones the subquery AST and invokes the generator with a \texttt{NULL} value injection on the projected column, generating a subquery record with a \texttt{NULL} join key.
    \item Combines the outer records and the null-injected subquery records into a unified dataset.
\end{enumerate}

For example, when evaluated on the dataset below where \texttt{student\_id} in \texttt{ExamScore} is \texttt{NULL}:
\begin{center}
\scriptsize
\begin{tabular}{cc}
  \begin{tabular}[t]{lc}
    \toprule
    \multicolumn{2}{c}{\textbf{Student (Outer)}} \\
    \texttt{id} & \texttt{name} \\
    \midrule
    1 & \textquotesingle Alice\textquotesingle \\
    \bottomrule
  \end{tabular}
  &
  \begin{tabular}[t]{lcc}
    \toprule
    \multicolumn{3}{c}{\textbf{ExamScore (Inner Subquery)}} \\
    \texttt{exam\_id} & \texttt{student\_id} & \texttt{score} \\
    \midrule
    101 & \texttt{NULL} & 85 \\
    \bottomrule
  \end{tabular}
\end{tabular}
\end{center}
The reference query \texttt{WHERE id NOT IN (SELECT student\_id FROM ExamScore)} evaluates to \texttt{UNKNOWN} and returns an empty result, whereas a mutant using \texttt{LEFT JOIN ... WHERE ExamScore.student\_id IS NULL} treats the row as unmatched and returns \textquotesingle Alice\textquotesingle{}, successfully killing the mutant.

\item \textbf{Alternative Column Attack Strategy:}
\label{sec:strategy_alt_col}
To kill column misselection mutants where a candidate query mistakenly filters on a type-compatible sibling attribute in the same table instead of the intended predicate column, this strategy transposes query literals across sibling columns:
\begin{enumerate}
    \item  Scans the table schema for alternative non-primary-key columns sharing data type compatibility with the predicate column, using attribute name similarity heuristics. (e.g., Digit Substitution such as phone1 $\to$ phone2, Affix Matching such as \texttt{ship\_city} $\to$ \texttt{bill\_city}, and similarly with prefixes/suffices such as \texttt{alt\_} and \texttt{\_other}).
    
    \item \textit{Literal Transposition Synthesis:} Generates a dataset using \texttt{value\_injections} that assigns the reference query's filter literal (e.g., \textquotesingle Paris\textquotesingle) to the sibling column while assigning a distinct non-matching value (\textquotesingle London\textquotesingle) to the original predicate column:
    \begin{center}
    \scriptsize
    \begin{tabular}{lccc}
      \toprule
      \multicolumn{4}{c}{\textbf{Orders}} \\
      \texttt{id} & \texttt{cust\_id} & \texttt{ship\_city} & \texttt{bill\_city} \\
      \midrule
      1 & 100 & \textquotesingle London\textquotesingle & \textquotesingle Paris\textquotesingle \\
      \bottomrule
    \end{tabular}
    \end{center}
\end{enumerate}

\end{itemize}

In total, the LASSO codebase implements $37$ distinct attack strategy classes. A breakdown of the top attack strategies and their corresponding refutation shares in evaluation is provided in Table~\ref{tab:attack_strategy_distribution}.

   \emph{Relation to Existing Techniques:} Selecting values near predicate boundaries is a known pattern from boundary value analysis~\cite{myers1979software} and domain testing principles~\cite{white1980domain}, which systematically select ON and OFF boundary points.  LASSO attack strategies include some techniques
   similar to those described in AGENDA~\cite{chays2004agenda, chays2004dissertation} such as filling duplicate values, forcing NULLs, and random selection of intermediate values from the intervals (LASSO chooses intermediate values deterministically).
   The alternative column attack strategy has some connections to fuzzy column matching used for schema matching.

\section{Evaluation Setup}
\label{sec:exec_test_setup}

Execution-Test (EX) is widely used in modern text-to-SQL benchmarks, such as BIRD, as the primary metric for determining semantic equivalence between two SQL queries. In standard EX, a gold query and a candidate predicted query are executed on an already populated database and considered non-equivalent (i.e. the prediction query is erroneous) if the query results are different.

However this approach ignores several subtle nuances that we address in our testing module \textsf{EX+}, an execution testing environment organized into two complementary phases: (i) pre-execution query normalization, and (ii) execution testing.

\subsection{Pre-Execution Query Normalization.}\label{sec:normalization}

Before queries are passed to test data generation tools or executed, \textsf{EX+} normalizes the input query AST using \texttt{sqlglot} to fix two sources of non-determinism.
\begin{itemize}
    \item \textbf{Time and Randomness Normalization}
    \begin{itemize}
        \item \textbf{Problem:} Queries invoking temporal functions (e.g., \texttt{NOW()}, \texttt{CURRENT\_TIMESTAMP}) or stochastic functions (e.g., \texttt{RANDOM()}) yield different outputs across execution calls, which can cause spurious mismatches on identical queries as well as on equivalent queries.
        \item \textbf{Check Implemented:} \textsf{EX+} transforms the query AST to rewrite temporal functions and string arguments (such as \texttt{'now'}) into a shared, frozen UTC timestamp, and seeds stochastic functions with a fixed seed.
    \end{itemize}

    \item \textbf{Sorting Tie-Breaking}
    \begin{itemize}
        \item \textbf{Problem:} When an \texttt{ORDER BY} clause sorts on attributes containing duplicate values (ties), SQL engines can return rows in physical orders depending on row layout and chosen execution plan. Consequently, two functionally equivalent queries can return identical rows in different sequences.  When combined with a LIMIT clause, identical or equivalent queries may potentially return different results. 
        \item \textbf{Check Implemented:} \textsf{EX+} augments the AST's \texttt{ORDER BY} clause by appending schema primary keys or unique columns (falling back to projected expressions) along with \texttt{NULLS LAST} to resolve sorting ties and produce stable, reproducible row ordering.
    \end{itemize}
\end{itemize}
Without the above normalization, execution testing may potentially report even identical but non-deterministic queries as non-equivalent.

\subsection{Execution Result Comparison}
\label{sec:result_comp}

During execution testing on generated datasets, \textsf{EX+} matches execution results and provides structured diagnostics through the following checks:
\begin{itemize}
    \item \textbf{Order and Multiset Sensitivity}:
 When the reference query contains an \texttt{ORDER BY} clause, \textsf{EX+} enforces strict list ordering. Otherwise, it compares results as unordered multisets using row frequency counters, accounting for discrepancies in duplicate tuple counts.

    \item \textbf{Permutation-Aware Column Matching:}
     Standard query output comparison requires exact positional column matching and equal column counts. However, text-to-SQL predictions may project columns in an alternate valid order (e.g., \texttt{(age, name)} instead of \texttt{(name, age)}) or project extra columns that might still fulfill the prompt.

     \textsf{EX+} supports a permutation-aware column matcher that evaluates whether any column permutation of the candidate query results matches that of the gold query results, and optionally accommodates extra columns.

     We note that the option did not have any significant impact on our benchmark queries, although it may be useful in other scenarios.

    \item \textbf{Strict Type Checking}:
    Standard Python comparators rely on loose equality where distinct primitive types are coerced (e.g., evaluating \texttt{1 == 1.0} or \texttt{1 == True} as true). While some applications may not require strict type checking, for others distinguishing data types is critical, such as for distinguishing boolean status flags from numeric codes, distinguishing a string containing numeric value from true numeric value, etc.
    
    \textsf{EX+} includes a configurable option that, when enabled, wraps each projected value with its runtime data type to prevent type coercion across integers, floats, strings, and booleans.

    \item \textbf{Structured Diagnostics:} Rather than returning an opaque binary pass/fail flag, \textsf{EX+} categorizes evaluation outcomes into fine-grained status codes (e.g. distinguishing true non-equivalent pairs, and non-equivalent pairs only due to differences in column ordering), failure reasons (e.g. row counts mismatch, values mismatch, etc.), and error classes (e.g. errors in query execution, schema loading, or dataset loading). 
    
    \item \textbf{Reproducible Counterexample Serialization}
       Whenever a counterexample dataset refutes equivalence, \textsf{EX+} automatically serializes a self-contained SQL script containing schema DDL, foreign-key safe data loading statements, both queries, and formatted execution outputs in block comments for immediate reproduction.
\end{itemize}

\subsection{Cross Checking Strategies}
\label{sec:cross_checking_strategies}

In our default, single-query data generation approach, test datasets are typically generated using only the reference query. These datasets are then used to evaluate candidate predicted queries. 

Datasets generated from a reference query may not trigger the specific logical pathways or boundary conditions of a predicted query. For instance, consider the following queries:

\begin{itemize}
    \item \textbf{Gold Query:} \texttt{SELECT name FROM students}
    \item \textbf{Predicted Query:} \texttt{SELECT name FROM students WHERE age <> 25}
\end{itemize}

Because the gold query lacks the \texttt{age <> 25} predicate, the datasets generated solely from the gold query may not generate records with \texttt{age = 25} that are needed to show non-equivalence of the two queries.


To address this, we utilize Two-Way Data Generation~\cite{chandra2015data,Chen2025ParSEvalPT}. If a candidate predicted query survives the datasets generated from the reference query, we use the predicted query to generate a supplementary set of ``Prediction Datasets.''  The comparison attack strategy operating on the predicted query will synthesize boundary data with \texttt{age = 25}, catching the above error.



We can discover more non-equivalent queries through cross-checking strategies. In practical settings, such as educational autograding or evaluating multiple text-to-SQL models on a benchmark, evaluators test multiple predictions $\mathcal{M} = \{Q_{pred}^{(1)}, \dots, Q_{pred}^{(m)}\}$ against a reference query $Q_{gold}$, and 
there are multiple queries one any given schema $\mathcal{S}$. Cross-checking pools and reuses test datasets across queries for a given schema, 
increasing the likelihood of exposing bugs, as demonstrated by experimentation (Table~\ref{tab:cross_checking_levels}).

\paragraph{Hierarchical Cross-Checking Scopes.}
Let an evaluation instance over database schema $\mathcal{S}$ consist of a reference query $Q_{gold}$ and candidate predictions $\mathcal{M} = \{Q_{pred}^{(1)}, \dots, Q_{pred}^{(m)}\}$, forming candidate query pairs $\mathcal{P} = \{(Q_{gold}, Q_{pred}) \mid Q_{pred} \in \mathcal{M}\}$. Across an entire benchmark, let $\mathcal{Q}_{\mathcal{S}}$ denote the set of all queries (both reference and predicted) that share schema $\mathcal{S}$.

For any query $Q$, let $\mathcal{D}(Q)$ denote the suite of test datasets synthesized by generating datasets from $Q$ (using LASSO's baseline and attack strategies). Unlike single-query testing where each query is evaluated solely on its own generated data, cross-checking evaluates candidate query pairs across four hierarchical levels:
\begin{enumerate}
    \item \textbf{Level 1 (Gold-Only):} 
    Datasets are generated solely from the reference query:
    \begin{equation}
        \mathcal{D}_1(p) = \mathcal{D}(Q_{gold})
    \end{equation}
    All candidate pairs $p \in \mathcal{P}$ are tested against $\mathcal{D}_1(p)$ and checked for differences in results.
    \item \textbf{Level 2 (Pairwise Cross-Checking):}
    Datasets are generated from both queries in the pair:
    \begin{equation}
        \mathcal{D}_2(p) = \mathcal{D}_1(p) \cup \mathcal{D}(Q_{pred})
    \end{equation}
    In an optimized setting, only the surviving (non-refuted) candidate query pairs after Level 1 are used to generate additional datasets based on the prediction query.
    \item \textbf{Level 3 (Instance-Level Sibling Pool):}
    Datasets are generated for all candidates (regardless of whether surviving or already killed) in $\mathcal{M}$ and pooled:
    \begin{equation}
        \mathcal{D}_3(p) = \mathcal{D}_1(p) \cup \bigcup_{j=1}^m \mathcal{D}(Q_{pred}^{(j)})
    \end{equation}
    \item \textbf{Level 4 (Cross-Instance Schema Pool):}
    Datasets generated across all queries $\mathcal{Q}_{\mathcal{S}}$ on schema $\mathcal{S}$ are pooled:
    \begin{equation}
        \mathcal{D}_4(p) = \bigcup_{Q \in \mathcal{Q}_{\mathcal{S}}} \mathcal{D}(Q)
    \end{equation}
\end{enumerate}


\section{Performance Evaluation}
\label{sec:performance}

To evaluate the effectiveness of LASSO, we compare it with recent test data generation tools on our unified benchmark.   All evaluation and benchmarking scripts were manually reviewed and checked using unit test cases, to ensure that the experimental results and timings are correct and reliable.

We note that LASSO and XData can be used in situations where only one query is available, e.g., a query written by a programmer or a text-to-SQL system, which needs to be validated by using test data.
The alternatives that we compare with, however, check for equivalence of a pair of queries.  To check for effectiveness of our data generation, we compare LASSO with the alternatives in terms of its effectiveness in detecting non-equivalence of query pairs.

\subsection{Unified Benchmark}
\label{sec:unified_benchmark}

We constructed a unified benchmark integrating query pairs and schemas across eight benchmark sources, which includes benchmarks used for originally testing each individual tool. We compared against: \textbf{BIRD}~\cite{bird} ($10,153$ pairs),
   \textbf{Calcite} (\textbf{Calc})~\cite{10.1145/3729303} ($221$ pairs),
   \textbf{Literature} (\textbf{Lit})~\cite{10.1145/3729303} ($51$ pairs),
   \textbf{Cubes}~\cite{10.1145/3729303} ($3,106$ pairs),
   \textbf{LeetCode} (\textbf{LC})~\cite{10.1145/3729303, Tremante2026SpotItVT} ($21,960$ pairs),
   \textbf{XDataBM} (\textbf{XBM})~\cite{xdata2011, Somwase2025TestDG} ($343$ pairs),
   \textbf{TPC-H}(\textbf{TPX})~\cite{Somwase2025TestDG} with manually created mutations by XData  ($198$ pairs),
   and \textbf{LargeJoin} (\textbf{LJ})~\cite{Somwase2025TestDG} ($10$ pairs).
In subsequent evaluation tables, these benchmark names are abbreviated accordingly.

We implemented some pre-processing and sanitization and used the normalized version of queries (generated as described in Section~\ref{sec:normalization}, and made SQLite/PostgreSQL compatible, as required, using SQLGlot) to prepare the inputs as required for various other tools.
We eliminated query pairs where we could trivially determine equivalence, syntactically invalid queries, duplicate prediction queries, etc. (The exact reasons for eliminations are available in our repo.)
Our final unified benchmark has  $36,042$ query pairs.\footnote{Two queries can also be proven non-equivalent if a dataset causes one query to execute successfully while triggering a runtime execution error in the other (for example, a subquery returning more than one row, a division by zero, or a numeric overflow). However, identifying such specific errors is not handled in our testing scripts and presented results only consider NEQ only when both queries execute successfully against the dataset and produce different output results.}

\subsection{Performance Results}
\label{sec:eval_results}

We compared LASSO  with four state-of-the-art test data generation tools: SpotIT~\cite{Tremante2026SpotItVT}, Polygon~\cite{10.1145/3729303}, ParSEval~\cite{Chen2025ParSEvalPT}, and XData~\cite{xdata2011}, as well as direct data generation using Gemini 3.8 Flash.  We also compared extensions LASSO-CC and XData-CC of LASSO and XData, which perform pairwise cross checking by generating data from both queries instead of from only one query.

We note that since the Bird, Cubes, LeetCode and Literature benchmarks are SQLite/MySQL based, wheareas XData works only on the PostgreSQL dialect of SQL, we did not run these benchmarks on XData.

To avoid cases where data generation takes a very long time on some queries, we set a timeout of 30 seconds for data generation for each query pair being compared (SpotIt, Polygon, ParSEval and Gemini).  For XData and LASSO, datasets are generated from one query and used against multiple candidate queries.  To make the comparison fair, we set the timeout for XData to 1 minute (for generating all datasets), but left it as 30 seconds for LASSO since LASSO was very fast anyway. 

\begin{table}
\centering
\caption{Refutations on the Unified Benchmark \textrm{(- means tool had many errors, * means we did not run the tool due to cost)}}
\label{tab:unified_results}
\resizebox{\columnwidth}{!}{%
\setlength{\tabcolsep}{3pt}
\begin{tabular}{l ccccccccc}
\toprule
& \multicolumn{8}{c}{\textbf{Benchmark}} & \\
\cmidrule(lr){2-9}
\textbf{Tool} & \textbf{BIRD} & \textbf{Cubes} & \textbf{LC} & \textbf{Lit} & \textbf{Calc} & \textbf{XBM} & \textbf{TPX} & \textbf{LJ} & \textbf{Total} \\
& ($10,153$) & ($3,106$) & ($21,960$) & ($51$) & ($221$) & ($343$) & ($198$) & ($10$) & ($36,042$) \\
\midrule
SpotIT & 1,048 & 576 & 1,414 & 15 & 1 & 0 & 11 & 3 & 3,068 \\
Polygon & 3,399 & 558 & 4,752 & 18 & 1 & 0 & 61 & 5 & 8,794 \\
ParSEval & 5,800 & 1,044 & 5,244 & 16 & 6 & 120 & 122 & 6 & 12,358 \\
XData & -- & -- & -- & -- & 2 & 302 & 135 & 4 & -- \\
XData-CC & -- & -- & -- & -- & 4 & \textbf{327} & 141 & 4 & -- \\
Gemini & * & * & * & * & * & * & \textbf{186} & \textbf{10} & * \\
LASSO & 6,255 & 2,320 & 6,984 & 19 & \textbf{9} & 318 & 159 & \textbf{10} & 16,074 \\
LASSO-CC & \textbf{6,685} & \textbf{2,466} & \textbf{11,279} & \textbf{22} & \textbf{9} & 324 & 162 & \textbf{10} & \textbf{20,957} \\
\bottomrule
\end{tabular}%
}
\end{table}

Table~\ref{tab:unified_results} presents the \textit{number of refutations per benchmark} across all $36,042$ query pairs.  For tools that support only a subset of benchmarks (such as XData on Calcite and XDataBM), unsupported benchmarks are marked with a dash (--), while * (for the case of Gemini) denotes that we did not run the experiment due to the high cost of running on thousands of query pairs.

It can be seen that LASSO performs significantly better than the alternatives across all benchmarks, except XDataBM where XData-CC performed marginally better, and LASSO-CC improves further on LASSO.
Gemini also performs very well, matching or surpassing LASSO for TPC-H (TPX) and LargeJoin (LJ) in terms of refutations, but as we discuss later, is much slower than LASSO.

\begin{table}
\centering
\caption{Refutations on Common Supported Queries}
\label{tab:common_supported_results}
\resizebox{\columnwidth}{!}{%
\setlength{\tabcolsep}{3pt}
\begin{tabular}{l ccccccccc}
\toprule
& \multicolumn{8}{c}{\textbf{Benchmark}} & \\
\cmidrule(lr){2-9}
\textbf{Tool} & \textbf{BIRD} & \textbf{Cubes} & \textbf{LC} & \textbf{Lit} & \textbf{Calc} & \textbf{XBM} & \textbf{TPX} & \textbf{LJ} & \textbf{Total} \\
& ($7,508$) & ($1,734$) & ($8,303$) & ($42$) & ($96$) & ($139$) & ($36$) & ($10$) & ($17,868$) \\
\midrule
SpotIT & 965 & 387 & 1,165 & 14 & 1 & 0 & 11 & 3 & 2,546 \\
Polygon & 3,223 & 402 & 2,089 & 18 & 0 & 0 & 35 & 5 & 5,772 \\
ParSEval & 4,111 & 442 & 2,536 & 16 & 4 & 100 & 14 & 6 & 7,229 \\
XData & -- & -- & -- & -- & 1 & 116 & 22 & 4 & -- \\
XData-CC & -- & -- & -- & -- & 3 & 130 & 28 & 4 & -- \\
Gemini & * & * & * & * & * & * & \textbf{36} & \textbf{10} & * \\
LASSO & 4,385 & 1,221 & 2,702 & 18 & \textbf{6} & 130 & 21 & \textbf{10} & 8,493 \\
LASSO-CC & \textbf{4,629} & \textbf{1,304} & \textbf{3,874} & 21 & \textbf{6} & \textbf{134} & 21 & \textbf{10} & \textbf{9,999} \\
\bottomrule
\end{tabular}%
}
\end{table}

\begin{table}
\centering
\caption{Refutations on Queries Excluding \texttt{ORDER BY} and \texttt{LIMIT}}
\label{tab:no_order_limit_results}
\resizebox{\columnwidth}{!}{%
\setlength{\tabcolsep}{3pt}
\begin{tabular}{l cccccccc}
\toprule
& \multicolumn{7}{c}{\textbf{Benchmark}} & \\
\cmidrule(lr){2-8}
\textbf{Tool} & \textbf{BIRD} & \textbf{LC} & \textbf{Lit} & \textbf{Calc} & \textbf{XBM} & \textbf{TPX} & \textbf{LJ} & \textbf{Total} \\
& ($7,825$) & ($15,118$) & ($51$) & ($207$) & ($343$) & ($32$) & ($10$) & ($23,586$) \\
\midrule
SpotIT & 624 & 1,177 & 15 & 0 & 0 & 0 & 3 & 1,819 \\
Polygon & 2,730 & 3,363 & 18 & 1 & 0 & 15 & 5 & 6,132 \\
ParSEval & 4,506 & 3,890 & 16 & 5 & 120 & 18 & 6 & 8,561 \\
XData & -- & -- & -- & 1 & 302 & 25 & 4 & -- \\
XData-CC & -- & -- & -- & 3 & \textbf{327} & 25 & 4 & -- \\
Gemini & * & * & * & * & * & \textbf{32} & \textbf{10} & * \\
LASSO & 4,681 & 5,340 & 19 & \textbf{6} & 318 & 19 & \textbf{10} & 10,393 \\
LASSO-CC & \textbf{4,944} & \textbf{7,952} & \textbf{22} & \textbf{6} & 324 & 19 & \textbf{10} & \textbf{13,277} \\
\bottomrule
\end{tabular}%
}
\end{table}

\eat{
\begin{table}[t]
\centering
\caption{Total Execution Time (seconds) ($36,042$ pairs)}
\label{tab:execution_time}
\resizebox{\columnwidth}{!}{%
\setlength{\tabcolsep}{3pt}
\begin{tabular}{l ccccccccc}
\toprule
& \multicolumn{8}{c}{\textbf{Benchmark}} & \\
\cmidrule(lr){2-9}
\textbf{Tool} & \textbf{BIRD} & \textbf{Cubes} & \textbf{LC} & \textbf{Lit} & \textbf{Calc} & \textbf{XBM} & \textbf{TPX} & \textbf{LJ} & \textbf{Total (s)} \\
\midrule
SpotIT & 11,432.9 & 1,218.6 & 76,041.0 & 110.9 & 98.6 & 200.6 & 1,296.6 & 141.7 & 90,541.0 \\
Polygon & \textbf{2,478.9} & 572.9 & 16,957.6 & 93.8 & 189.1 & 85.7 & 92.1 & 2.6 & 20,472.8 \\
ParSEval & 24,811.0 & 2,203.8 & 17,021.6 & 184.8 & 246.6 & 257.5 & 2,032.6 & 135.5 & 46,893.4 \\
XData & -- & -- & -- & -- & 1,085.8 & 289.8 & 359.6 & 443.9 & -- \\
XData-CC & -- & -- & -- & -- & 1,861.1 & 448.8 & 819.4 & 450.7 & -- \\
Gemini & * & * & * & * & * & * & 1364.1 & 59.5 & * \\
LASSO & 2,695.1 & \textbf{438.1} & \textbf{34.1} & \textbf{38.7} & \textbf{12.8} & \textbf{5.9} & \textbf{34.6} & \textbf{1.0} & \textbf{3,260.3} \\
LASSO-CC & 11,714.4 & 787.2 & 3,473.1 & 73.2 & 25.5 & 8.3 & 432.6 & \textbf{1.0} & 16,515.2 \\
\bottomrule
\end{tabular}%
}
\end{table}
}
\begin{table}
\centering
\caption{Average Time per Refutation (seconds) ($36,042$ pairs)}
\label{tab:time_per_refutation}
\resizebox{\columnwidth}{!}{%
\setlength{\tabcolsep}{3pt}
\begin{tabular}{l ccccccccc}
\toprule
& \multicolumn{8}{c}{\textbf{Benchmark}} & \\
\cmidrule(lr){2-9}
\textbf{Tool} & \textbf{BIRD} & \textbf{Cubes} & \textbf{LC} & \textbf{Lit} & \textbf{Calc} & \textbf{XBM} & \textbf{TPX} & \textbf{LJ} & \textbf{Overall} \\
\midrule
SpotIT & 10.91 & 2.12 & 53.78 & 7.39 & 98.64 & -- & 117.87 & 47.25 & 29.51 \\
Polygon & 0.73 & 1.03 & 3.57 & 5.21 & 189.10 & -- & 1.51 & 0.53 & 2.33 \\
ParSEval & 4.28 & 2.11 & 3.25 & 11.55 & 41.10 & 2.15 & 16.66 & 22.59 & 3.79 \\
XData & -- & -- & -- & -- & 542.91 & 0.96 & 2.66 & 110.97 & -- \\
XData-CC & -- & -- & -- & -- & 465.27 & 1.37 & 5.81 & 112.68 & -- \\
Gemini & * & * & * & * & * & * & 7.33 & 5.95 & * \\
LASSO & \textbf{0.43} & \textbf{0.19} & \textbf{<0.01} & \textbf{2.04} & \textbf{1.42} & \textbf{0.02} & \textbf{0.22} & \textbf{0.10} & \textbf{0.20} \\
LASSO-CC & 1.75 & 0.32 & 0.31 & 3.33 & 2.83 & 0.03 & 2.67 & \textbf{0.10} & 0.79 \\
\bottomrule
\end{tabular}%
}
\end{table}

Table~\ref{tab:common_supported_results} reports the \textit{refutation counts on the commonly supported subset} ($17,868$ query pairs).  Here, if at least one of the tools encounters an error during data generation, 
for e.g. because of SQL operators that the tool is not designed for or lacks support, or parser exceptions, etc., then the corresponding benchmark instance is excluded from results of all the tools.  Even on this benchmark, we see that LASSO and LASSO-CC outperform the alternatives overall.  ParSEval and Polygon close the gap here, implying that there are many queries they don't support, which LASSO supports, although on queries that they do support, they do well at refutation.

Table~\ref{tab:no_order_limit_results} presents refutation counts on the $23,586$ query pairs where neither the reference query nor the candidate prediction contains an \texttt{ORDER BY} or \texttt{LIMIT} clause, since some tools may not handle ordering.
Again, LASSO and LASSO-CC were substantially better than the alternatives, except on XDataBM where XData-CC was marginally better.
We note that the Cubes benchmark is eliminated here because all of the queries have an ORDER BY clause.

\textit{Execution time:}
Table~\ref{tab:time_per_refutation} reports the \textit{average time per refutation} (in seconds) achieved across each benchmark, as well as overall. For each tool and benchmark, this metric is computed as the total tool execution time spent on that benchmark divided by the number of confirmed refutations achieved ($\text{Execution Time} / \text{Refutations}$). 
The recorded time accounts only for tool execution time, and does not include the external time used for execution testing (loading datasets and executing queries). 

It can be seen that LASSO has the lowest execution time for each benchmark, and overall.



Solver-based tools (such as SpotIT, ParSEval, and XData) frequently time out or require substantial generation time when solving constraints for large or complex queries.  For the Large Join query benchmark (which has queries with 2, 4, 7, 8, and 10 relations being joined, with two instances of each size),  SpotIt and ParSEval timed out on 4 of 10 LJ queries, while XData-CC timed out on 6 of them.  We note that although Polygon had a relatively small  execution time on LJ, it failed to refute 5 out of 10 query pairs.

Direct test data synthesis using an LLM (Gemini 3.8 Flash) 
successfully generates refutation datasets for all $10$ LJ query pairs, but takes a total of $59.5$s, with 16K output and 6.3K input tokens (cost \$0.053).

In contrast, LASSO leveraging fast heuristic, successfully generates datasets on all $10$ queries in just $1.0$ second, and all 10 query pairs were refuted.

\noindent \textit{Effectiveness on Complex Queries/Unseen Schemas:}
Does LASSO generalize across schemas and complex queries that were not used during framework development is an important question.  We note that the LJ and  TPC benchmarks were not used during development, and thus unseen by the code base. 

We also evaluated LASSO 
on the BEAVER enterprise data warehouse (\texttt{dw}) suite~\cite{Chen2024BEAVERAE}, which has relatively complex queries, and which also we did not use during development.  Results are shown on 
10 queries from the dw schema.\footnote{
Getting results on the full set of schemas and queries after making them SQLite compatible is ongoing work.}

The results are summarized in 
Table~\ref{tab:beaver_results}, ParSEval produces non-empty datasets for only $5$ of the $10$ queries ($50.0\%$), timing out ($>30$s) on the other $5$ queries ($206.3$s total time) due to SMT constraint solver bottlenecks on complex CTEs. Direct LLM dataset generation using Gemini 3.8 Flash (\texttt{gemini-3.8-flash}) successfully synthesizes non-empty datasets for all $10$ queries ($100.0\%$), taking $43.4$s (15K input and 9k output tokens, total cost \$0.045) using API calls. In comparison, LASSO generates valid non-empty test datasets for $9$ of the $10$ queries ($90.0\%$) in just $0.16$s.

For TPX benchmark on Gemini, across 198 query pairs, Gemini took approx.\ 300k input and 300k output tokens for a total cost of \$1.35.

\begin{table}[t]
\centering
\caption{Data Generation on BEAVER Benchmark ($10$ Queries, \texttt{dw} Schema)}
\label{tab:beaver_results}
\small
\setlength{\tabcolsep}{6pt}
\begin{tabular}{lccc}
\toprule
\textbf{Tool} & \textbf{Non-Empty} & \textbf{TIMEOUT} & \textbf{Total Time (s)} \\
\midrule
ParSEval & 5 / 10  & 5 & 206.3 \\
Gemini 3.8 Flash & \textbf{10 / 10 } & \textbf{0} & 43.4  \\
LASSO & 9 / 10  & \textbf{0} & \textbf{0.16} \\
\bottomrule
\end{tabular}
\end{table} 


\noindent \textit{Generated Dataset Statistics:}
For each data generation call, LASSO generates a mean of $20.1$ datasets (median: $16.0$, standard deviation: $19.5$).   Furthermore, across all refutations, the number of tuples per dataset generated by LASSO has a mean of $5.3$ rows (median: $3.0$ rows, standard deviation: $31.5$ rows). The vast majority of generated counterexamples contain only a few records, they remain concise and easy to inspect for debugging.

Interestingly, one of the benchmark queries (Literature Q46) with a constraint of count(FID) $> 1000$ required at least 1000 rows in the dataset.  Only LASSO was able to successfully generate the dataset.

\begin{table}[t]
\centering
\caption{Effectiveness of Top 20 Individual Attack Strategies}
\label{tab:attack_strategy_distribution}
\small
\resizebox{2.5in}{!}{%
\setlength{\tabcolsep}{4pt}
\begin{tabular}{clr}
\toprule
\textbf{\#} & \textbf{Attack Strategy} & \textbf{\% of Refutations} \\
\midrule
1 & \texttt{TableAttackStrategy} & 8.32\% \\
2 & \texttt{BaseDatasetStrategy} & 7.98\% \\
3 & \texttt{JoinAttackStrategy} & 7.77\% \\
4 & \texttt{ComparisonAttackStrategy} & 6.99\% \\
5 & \texttt{GroupByAttackStrategy} & 6.45\% \\
6 & \texttt{AggregateNullAttackStrategy} & 6.37\% \\
7 & \texttt{MissingPredicateAttackStrategy} & 6.26\% \\
8 & \texttt{JoinCardinalityAttackStrategy} & 5.46\% \\
9 & \texttt{AggregateFunctionAttackStrategy} & 4.46\% \\
10 & \texttt{OrderingAttackStrategy} & 3.97\% \\
11 & \texttt{DuplicatesAttackStrategy} & 3.82\% \\
12 & \texttt{NullableJoinAttackStrategy} & 3.57\% \\
13 & \texttt{StringLiteralAttackStrategy} & 3.54\% \\
14 & \texttt{RedundantDistinctAttackStrategy} & 3.19\% \\
15 & \texttt{DistinctAttackStrategy} & 2.52\% \\
16 & \texttt{StringLiteralCaseAttackStrategy} & 2.27\% \\
17 & \texttt{JoinTypeAttackStrategy} & 1.88\% \\
18 & \texttt{DivisionAttackStrategy} & 1.64\% \\
19 & \texttt{AlternativeColumnAttackStrategy} & 1.50\% \\
20 & \texttt{AggregateMissingDistinctAttackStrategy} & 1.43\% \\
\bottomrule
\end{tabular}%
}
\end{table}

\textit{LASSO Attack Strategy Effectiveness:}
To analyze the effectiveness of different LASSO attack strategies,
we generated datasets from the  gold queries, with each dataset associated with the strategy that generated it. 
We then executed all query pairs on these datasets.

Table~\ref{tab:attack_strategy_distribution} presents the effectiveness breakdown of the top 20 attack strategies (sorted in descending order of refutations), reporting the percentage share of total strategy kills contributed by each strategy. 

The most effective strategy,   \textit{TableAttackStrategy} injects parent rows that have no matching child rows, to kill mutants that substitute a table by another, omit a table, or add spurious outer joins to a parent table.
The \textit{BaseDatasetStrategy}  generates datasets to provide non-empty results on the query, killing all mutants that give empty results. \textit{JoinAttackStrategy} generates multiple datasets having one empty table at a time to detect SQL queries that use the wrong JOIN type and incorrectly include or exclude rows. The \textit{DivisionAttackStrategy}  generates datasets to produce 0, negative, or NULL values for division denominators. A detailed explanation of each of the attack strategies is available in our repo.

\begin{table}[t]
\centering
\caption{Refutation Origin in LASSO-CC}
\label{tab:refutation_source_breakdown}
\small
\setlength{\tabcolsep}{6pt}
\begin{tabular}{lcccc}
\toprule
\textbf{Benchmark} & \textbf{Total} & \textbf{Gold} & \textbf{Pred (CC)} & \textbf{Gain} \\
\midrule
BIRD & $6,685$ & $6,255$ & $430$ & $+6.9\%$ \\
Cubes & $2,466$ & $2,320$ & $146$ & $+6.3\%$ \\
LeetCode & $11,279$ & $6,984$ & $4,295$ & $+61.5\%$ \\
Literature & $22$ & $19$ & $3$ & $+15.8\%$ \\
Calcite & $9$ & $9$ & $0$ & $+0.0\%$ \\
XDataBM & $324$ & $318$ & $6$ & $+1.9\%$ \\
TPC-H & $162$ & $159$ & $3$ & $+1.9\%$ \\
LargeJoin & $10$ & $10$ & $0$ & $+0.0\%$ \\
\midrule
\textbf{Overall} & \textbf{20,957} & \textbf{16,074} & \textbf{4,883} & \textbf{+30.4\%} \\
\bottomrule
\end{tabular}

\end{table}

\noindent \textit{Performance of Cross-Checking Strategies}:
\label{sec:cross_checking_study}
To evaluate the \textit{effectiveness of pair-wise cross-checking} (LASSO-CC), we compared the number of datasets generated from gold queries that could refute equivalence, versus the incremental benefit of generating datasets from prediction queries when the gold query generated datasets failed to refute equivalence.  
Table~\ref{tab:refutation_source_breakdown} shows these results across each benchmark suite.
Overall, the incremental gain due to cross-checking was over $60\%$ for LeetCode, and $30\%$ overall although less for other benchmarks.

To evaluate the impact of all the four cross-checking strategies (Section~\ref{sec:cross_checking_strategies}), we tested $100$ BIRD benchmark instances spanning $11$ database schemas sequentially from Level 1 to Level 4, recording the first level at which each non-equivalent mutant is killed and the execution runtime. As observed from Table~\ref{tab:cross_checking_levels}, using the generated test datasets for queries other than the reference query makes it possible to discover more non-equivalent query pairs, though at the expense of higher evaluation time due to the substantially larger pool of test datasets evaluated.

\begin{table}[t]
\centering
\caption{Refutation Performance Across Cross-Checking Levels ($100$ BIRD Instances, $664$ Query Pairs Across $11$ Schemas)}
\label{tab:cross_checking_levels}
\small
\resizebox{\columnwidth}{!}{%
\setlength{\tabcolsep}{2pt}
\begin{tabular}{ll cccc}
\toprule
\textbf{Level} & \textbf{Dataset Scope} & \textbf{Killed} & \textbf{Frac.} & \textbf{Marg. Gain} & \textbf{Time (s)} \\
\midrule
Level 1 & Gold-Only ($\mathcal{D}_1$) & 411 & 61.9\% & -- & 1.80 \\
Level 2 & Pairwise Gold + Pred ($\mathcal{D}_2$) & 431 & 64.9\% & +20 (+3.0\%) & 2.89 \\
Level 3 & Instance Pool ($\mathcal{D}_3$) & 436 & 65.7\% & +5 (+0.8\%) & 5.08 \\
Level 4 & Schema Pool ($\mathcal{D}_4$) & \textbf{439} & \textbf{66.1\%} & +3 (+0.5\%) & 55.33 \\
\midrule
\textbf{Net Gain} & \textbf{Full Cross-Checking } & \textbf{439} & \textbf{66.1\%} & \textbf{+28 (+4.2\%)} & \textbf{65.09} \\
\bottomrule
\end{tabular}%
}
\end{table}

\section{Conclusion}
\label{sec:conclusion}

In this paper we described how we built the LASSO test data generation system using agentic coding.  We demonstrated that it is possible to generate test data using case-specific techniques with low execution time, and yet refute more query pairs than existing techniques.  The approach we used for creating LASSO agentically can be useful for other similar projects in future.


An important area of future work is to test and extend LASSO to handle queries of high complexity, beyond what we have in the current benchmarks.

Another area of future work in the context of text-to-SQL is to explore how to use generated test datasets and user feedback on correctness of results to guide the LLM to correct errors in the generated SQL.
Expanding attack strategies currently requires human oversight, since agents are observed to violate rules such as avoiding hard-coded constants, random data generation etc.  
How to use agents to detect and prevent such violations is an area of future work.  Some queries result in generation of a large number of datasets.  Minimizing the number of datasets is another area of future work.

\bibliographystyle{ACM-Reference-Format}
\bibliography{sample-base}

@String{Computing = "Computing" }

@String{Computer = "{IEEE} Computer" }

@String{Springer = "Springer-Verlag" }

@inproceedings{xdata2011,
  author={Shah, Shetal and Sudarshan, S. and Kajbaje, Suhas and Patidar, Sandeep and Gupta, Bhanu Pratap and Vira, Devang},
  booktitle={2011 IEEE 27th International Conference on Data Engineering}, 
  title={Generating test data for killing SQL mutants: A constraint-based approach}, 
  year={2011},
  volume={},
  number={},
  pages={1175-1186},
  doi={10.1109/ICDE.2011.5767876},
  url       = {https://doi.org/10.1109/ICDE.2011.5767876}
}

@article{chandra2015data,
author = {Chandra, Bikash and Chawda, Bhupesh and Kar, Biplab and Reddy, K. V. and Shah, Shetal and Sudarshan, S.},
title = {Data generation for testing and grading SQL queries},
year = {2015},
issue_date = {December  2015},
publisher = {Springer-Verlag},
address = {Berlin, Heidelberg},
volume = {24},
number = {6},
issn = {1066-8888},
url = {https://doi.org/10.1007/s00778-015-0395-0},
doi = {10.1007/s00778-015-0395-0},
journal = {The VLDB Journal},
month = dec,
pages = {731–755},
numpages = {25}
}

@book{myers1979software,
 author = {Myers, Glenford J. and Sandler, Corey and Badgett, Tom},
title = {The Art of Software Testing},
year = {2011},
isbn = {1118031962},
publisher = {Wiley Publishing},
edition = {3rd},
  url       = {https://dl.acm.org/doi/10.5555/2161638}
}

@inproceedings{veanes2010qex,
author="Veanes, Margus
and Tillmann, Nikolai
and de Halleux, Jonathan",
editor="Clarke, Edmund M.
and Voronkov, Andrei",
title="Qex: Symbolic SQL Query Explorer",
booktitle="Logic for Programming, Artificial Intelligence, and Reasoning",
year="2010",
publisher="Springer Berlin Heidelberg",
address="Berlin, Heidelberg",
pages="425--446",
isbn="978-3-642-17511-4",
  doi       = {10.1007/978-3-642-17511-4_24},
  url       = {https://doi.org/10.1007/978-3-642-17511-4_24}
}

@inproceedings{binnig2007reverse,
 author={Binnig, Carsten and Kossmann, Donald and Lo, Eric},
  booktitle={2007 IEEE 23rd International Conference on Data Engineering}, 
  title={Reverse Query Processing}, 
  year={2007},
  volume={},
  number={},
  pages={506-515},
  doi={10.1109/ICDE.2007.367896},
  ISSN={2375-026X},
  month={April},
  url       = {https://doi.org/10.1109/ICDE.2007.367896}
}

@inproceedings{binnig2007qagen,
 title           = "{QAGen}",
  booktitle       = "Proceedings of the 2007 {ACM} {SIGMOD} international
                     conference on Management of data",
  author          = "Binnig, Carsten and Kossmann, Donald and Lo, Eric and
                     {\"O}zsu, M Tamer",
  publisher       = "ACM",
  month           =  jun,
  year            =  2007,
  address         = "New York, NY, USA",
  conference      = "SIGMOD/PODS07: International Conference on Management of
                     Data",
  location        = "Beijing China",
  url       = {https://doi.org/10.1145/1247480.1247520}
}

@inproceedings{gupta2010xdata,
  author={Gupta, Bhanu Pratap and Vira, Devang and Sudarshan, S.},
  booktitle={2010 IEEE 26th International Conference on Data Engineering (ICDE 2010)}, 
  title={X-data: Generating test data for killing SQL mutants}, 
  year={2010},
  volume={},
  number={},
  pages={876-879},
  doi={10.1109/ICDE.2010.5447862},
  url       = {https://doi.org/10.1109/ICDE.2010.5447862}
}

@article{Chen2025ParSEvalPT,
author = {Chen, Chunyu and Miao, Zhengjie and Zhang, Yong and Wang, Jiannan},
title = {ParSEval: Plan-Aware Test Database Generation for SQL Equivalence Evaluation},
year = {2025},
issue_date = {July 2025},
publisher = {VLDB Endowment},
volume = {18},
number = {11},
issn = {2150-8097},
url = {https://doi.org/10.14778/3749646.3749727},
doi = {10.14778/3749646.3749727},
journal = {Proc. VLDB Endow.},
month = jul,
pages = {4750–4762},
numpages = {13}
}

@article{Tremante2026SpotItVT,
  author    = {Andrew Tremante and Yang He and Rocky Klopfenstein and Yuepeng Wang and Nina Narodytska and Haoze Wu},
  title     = {{SpotIt+}: Verification-based {Text-to-SQL} Evaluation with Database Constraints},
  journal   = {arXiv preprint arXiv:2603.04334},
  year      = {2026},
  doi       = {10.48550/arXiv.2603.04334},
  url       = {https://doi.org/10.48550/arXiv.2603.04334}
}

@article{10.1145/3729303,
   title     = "Polygon: Symbolic reasoning for {SQL} using conflict-driven
               under-approximation search",
  author    = "Zhao, Pinhan and Wang, Yuepeng and Wang, Xinyu",
  journal   = "Proc. ACM Program. Lang.",
  publisher = "Association for Computing Machinery (ACM)",
  volume    =  9,
  number    = "PLDI",
  pages     = "1315--1340",
  month     =  jun,
  year      =  2025,
  copyright = "http://www.acm.org/publications/policies/copyright\_policy\#Background",
  language  = "en",
  url       = {https://doi.org/10.1145/3729303}
}

@article{Somwase2025TestDG,
author = {Somwase, Sunanda and Das, Parismita and Sudarshan, S.},
title = {Test Data Generation for Complex SQL Queries},
year = {2025},
issue_date = {December 2025},
publisher = {Association for Computing Machinery},
address = {New York, NY, USA},
volume = {3},
number = {6},
url = {https://doi.org/10.1145/3769832},
doi = {10.1145/3769832},
journal = {Proc. ACM Manag. Data},
month = dec,
articleno = {367},
numpages = {26}

}

@inproceedings{gray1994quickly,
 author = {Gray, Jim and Sundaresan, Prakash and Englert, Susanne and Baclawski, Ken and Weinberger, Peter J.},
title = {Quickly generating billion-record synthetic databases},
year = {1994},
isbn = {0897916395},
publisher = {Association for Computing Machinery},
address = {New York, NY, USA},
url = {https://doi.org/10.1145/191839.191886},
doi = {10.1145/191839.191886},
booktitle = {Proceedings of the 1994 ACM SIGMOD International Conference on Management of Data},
pages = {243–252},
numpages = {10},
location = {Minneapolis, Minnesota, USA},
series = {SIGMOD '94}
}

@inproceedings{bruno2005flexible,
 author = {Bruno, Nicolas and Chaudhuri, Surajit},
title = {Flexible database generators},
year = {2005},
isbn = {1595931546},
publisher = {VLDB Endowment},
booktitle = {Proceedings of the 31st International Conference on Very Large Data Bases},
pages = {1097–1107},
numpages = {11},
location = {Trondheim, Norway},
series = {VLDB '05},
  url       = {http://www.vldb.org/conf/2005/papers/p1097-bruno.pdf}
}

@article{chays2004agenda,
  author = {Chays, David and Deng, Yuetang and Frankl, Phyllis G. and Dan, Saikat and Vokolos, Filippos I. and Weyuker, Elaine J.},
title = {An AGENDA for testing relational database applications},
journal = {Software Testing, Verification and Reliability},
volume = {14},
number = {1},
pages = {17-44},
doi = {https://doi.org/10.1002/stvr.286},
url = {https://onlinelibrary.wiley.com/doi/abs/10.1002/stvr.286},
eprint = {https://onlinelibrary.wiley.com/doi/pdf/10.1002/stvr.286},
year = {2004}
}

@phdthesis{chays2004dissertation,
  author    = {David Chays},
  title     = {Test Data Generation for Relational Database Applications},
  school    = {Polytechnic University},
  year      = {2004},
  address   = {Brooklyn, New York},
  note      = {Department of Computer and Information Science, TR-CIS-2005-01},
  url       = {https://cs.ecu.edu/gudivada/research/papers/Database-Test-Data-Generation.pdf}
}

@article{white1980domain,
  author    = {Lee J. White and Edward I. Cohen},
  title     = {A Domain Strategy for Computer Program Testing},
  journal   = {IEEE Transactions on Software Engineering},
  volume    = {SE-6},
  number    = {3},
  pages     = {247--257},
  year      = {1980},
  publisher = {IEEE},
  doi       = {10.1109/TSE.1980.234486},
  url       = {https://doi.org/10.1109/TSE.1980.234486}
}

@inproceedings{bird,
  author = {Li, Jinyang and Hui, Binyuan and Qu, Ge and Yang, Jiaxi and Li, Binhua and Li, Bowen and Wang, Bailin and Qin, Bowen and Geng, Ruiying and Huo, Nan and Zhou, Xuanhe and Ma, Chenhao and Li, Guoliang and Chang, Kevin C.C. and Huang, Fei and Cheng, Reynold and Li, Yongbin},
  title = {Can LLM already serve as a database interface? a big bench for large-scale database grounded text-to-SQLs},
  booktitle = {Proceedings of the 37th International Conference on Neural Information Processing Systems},
  year = {2023},
  publisher = {Curran Associates Inc.},
  address = {Red Hook, NY, USA},
  articleno = {1835},
  numpages = {28},
  location = {New Orleans, LA, USA},
  series = {NIPS '23},
  url = {https://dl.acm.org/doi/abs/10.5555/3666122.3667957}
}

@inproceedings{Liu2025NL2SQLBUGsAB,
author = {Liu, Xinyu and Shen, Shuyu and Li, Boyan and Tang, Nan and Luo, Yuyu},
title = {NL2SQL-BUGs: A Benchmark for Detecting Semantic Errors in NL2SQL Translation},
year = {2025},
isbn = {9798400714542},
publisher = {Association for Computing Machinery},
address = {New York, NY, USA},
url = {https://doi.org/10.1145/3711896.3737427},
doi = {10.1145/3711896.3737427},
booktitle = {Proceedings of the 31st ACM SIGKDD Conference on Knowledge Discovery and Data Mining V.2},
pages = {5662–5673},
numpages = {12},
location = {Toronto ON, Canada},
series = {KDD '25}}

@inproceedings{zhong-etal-2020-semantic,
  title     = {Semantic Evaluation for {T}ext-to-{SQL} with Distilled Test Suites},
  author    = {Zhong, Ruiqi and Yu, Tao and Klein, Dan},
  booktitle = {Proceedings of the 2020 Conference on Empirical Methods in Natural Language Processing (EMNLP)},
  pages     = {396--411},
  year      = {2020},
  publisher = {Association for Computational Linguistics},
  doi       = {10.18653/v1/2020.emnlp-main.29},
  url       = {https://doi.org/10.18653/v1/2020.emnlp-main.29}
}

@inproceedings{zhao2025llmsqlsolver,
  author    = {Fuheng Zhao and Jiayue Chen and Lawrence Lim and Ishtiyaque Ahmad and Divyakant Agrawal and Amr {El Abbadi}},
  title     = {{LLM-SQL-Solver}: Can {LLMs} Determine {SQL} Equivalence?},
  booktitle = {Proceedings of the 2025 IEEE International Conference on Big Data (BigData)},
  pages     = {1887--1894},
  year      = {2025},
  publisher = {IEEE},
  doi       = {10.1109/BigData66926.2025.11401595},
  url       = {https://doi.org/10.1109/BigData66926.2025.11401595}
}

@inproceedings{swami1994estimation,
  author    = {Arun N. Swami and K. Bernhard Schiefer},
  title     = {On the Estimation of Join Result Sizes},
  booktitle = {Advances in Database Technology --- {EDBT} '94, 4th International Conference on Extending Database Technology},
  series    = {Lecture Notes in Computer Science},
  volume    = {779},
  pages     = {287--300},
  publisher = {Springer, Berlin, Heidelberg},
  year      = {1994},
  doi       = {10.1007/3-540-57818-8_58},
  url       = {https://doi.org/10.1007/3-540-57818-8_58},
  isbn      = {978-3-540-57818-5}
}

@article{liu-etal-2024-lost,
    title = "Lost in the Middle: How Language Models Use Long Contexts",
    author = "Liu, Nelson F.  and
      Lin, Kevin  and
      Hewitt, John  and
      Paranjape, Ashwin  and
      Bevilacqua, Michele  and
      Petroni, Fabio  and
      Liang, Percy",
    journal = "Transactions of the Association for Computational Linguistics",
    volume = "12",
    year = "2024",
    address = "Cambridge, MA",
    publisher = "MIT Press",
    url = "https://aclanthology.org/2024.tacl-1.9/",
    doi = "10.1162/tacl_a_00638",
    pages = "157--173"
}

@article{Chen2024BEAVERAE,
  title={{BEAVER}: An Enterprise Benchmark for Text-to-{SQL}},
  author={Peter Baile Chen and Weiyue Li and Fabian Wenz and Yi Zhang and Yi Zhang and Nesime Tatbul and Michael J. Cafarella and {\c{C}}a{\u{g}}atay Demiralp and Michael Stonebraker},
  journal={ArXiv},
  year={2024},
  volume={abs/2409.02038},
  url={https://api.semanticscholar.org/CorpusID:272367292}
}

@article{verieql,
author = {He, Yang and Zhao, Pinhan and Wang, Xinyu and Wang, Yuepeng},
title = {VeriEQL: Bounded Equivalence Verification for Complex SQL Queries with Integrity Constraints},
year = {2024},
issue_date = {April 2024},
publisher = {Association for Computing Machinery},
address = {New York, NY, USA},
volume = {8},
number = {OOPSLA1},
url = {https://doi.org/10.1145/3649849},
doi = {10.1145/3649849},
journal = {Proc. ACM Program. Lang.},
month = apr,
articleno = {132},
numpages = {29}
}

@inproceedings{cosette,
  author       = {Shumo Chu and
                  Chenglong Wang and
                  Konstantin Weitz and
                  Alvin Cheung},
  title        = {Cosette: An Automated Prover for {SQL}},
  booktitle    = {8th Biennial Conference on Innovative Data Systems Research, {CIDR}
                  2017, Chaminade, CA, USA, January 8-11, 2017, Online Proceedings},
  publisher    = {www.cidrdb.org},
  year         = {2017},
  url          = {http://cidrdb.org/cidr2017/papers/p51-chu-cidr17.pdf},
  bibsource    = {dblp computer science bibliography, https://dblp.org}
}

@inproceedings{spes,
  author       = {Qi Zhou and
                  Joy Arulraj and
                  Shamkant B. Navathe and
                  William Harris and
                  Jinpeng Wu},
  title        = {{SPES:} {A} Symbolic Approach to Proving Query Equivalence Under Bag
                  Semantics},
  booktitle    = {38th {IEEE} International Conference on Data Engineering, {ICDE} 2022,
                  Kuala Lumpur, Malaysia, May 9-12, 2022},
  pages        = {2735--2748},
  publisher    = {{IEEE}},
  year         = {2022},
  url          = {https://doi.org/10.1109/ICDE53745.2022.00250},
  doi          = {10.1109/ICDE53745.2022.00250},
  bibsource    = {dblp computer science bibliography, https://dblp.org}
}

@article{QED,
author = {Wang, Shuxian and Pan, Sicheng and Cheung, Alvin},
title = {QED: A Powerful Query Equivalence Decider for SQL},
year = {2024},
issue_date = {July 2024},
publisher = {VLDB Endowment},
volume = {17},
number = {11},
issn = {2150-8097},
url = {https://doi.org/10.14778/3681954.3682024},
doi = {10.14778/3681954.3682024},
journal = {Proc. VLDB Endow.},
month = jul,
pages = {3602–3614},
numpages = {13}
}
\twocolumn
\end{document}